\documentclass[aps,prb,singlecolumn,superscriptaddress,showpacs]{revtex4-1}
\usepackage{bm}
\usepackage[dvips]{graphicx}
\usepackage{amsmath,amssymb}
\usepackage{epsfig}
\usepackage{color}
\usepackage{dcolumn}

\newcommand{\ketg}{\mbox{$|g\rangle $}}

\begin{document}
\title{Coherent control theory and experiment of optical phonons in diamond} 

\author{Hiroya Sasaki}
\affiliation{Laboratory for Materials and Structures, Institute of Innovative Research, Tokyo Institute of Technology, 4259 Nagatsuta, Yokohama 226-8503, Japan}

\author{Riho Tanaka}
\affiliation{Laboratory for Materials and Structures, Institute of Innovative Research, Tokyo Institute of Technology, 4259 Nagatsuta, Yokohama 226-8503, Japan}

\author{Yasuaki Okano}
\email[]{yokano@ims.ac.jp}
\affiliation{Center for Mesoscopic Sciences, Institute for Molecular Science, National Institutes of Natural Sciences, 38 Nishigo-Naka, Myodaiji, Okazaki 444-8585, Japan}

\author{Fujio Minami}
\affiliation{Laboratory for Materials and Structures, Institute of Innovative Research, Tokyo Institute of Technology, 4259 Nagatsuta, Yokohama 226-8503, Japan}
\affiliation{Department of Physics, Graduate School and Faculty of Engineering, Yokohama National University, 79-5 Tokiwadai, Hodogaya, Yokohama 240-8501 Japan}

\author{Yosuke Kayanuma}
\affiliation{Laboratory for Materials and Structures, Institute of Innovative Research, Tokyo Institute of Technology, 4259 Nagatsuta, Yokohama 226-8503, Japan}
\affiliation{Graduate School of Sciences, Osaka Prefecture University, 1-1 Gakuen-cho, Sakai, Osaka, 599-8531 Japan}

\author{Yutaka Shikano}
\email{yutaka.shikano@keio.jp}
\affiliation{Quantum Computing Center, Keio University, 3-14-1 Hiyoshi, Kohoku, Yokohama 223-8522, Japan}
\affiliation{Research Center for Advanced Science and Technology (RCAST), The University of Tokyo, 4-6-1 Komaba, Meguro, Tokyo 153-8904, Japan}
\affiliation{Institute for Quantum Studies, Chapman University, 1 University Dr., Orange, California 92866, USA}
\affiliation{Research Center of Integrative Molecular Systems (CIMoS), Institute for Molecular Science, National Institutes of Natural Sciences, 38 Nishigo-Naka, Myodaiji, Okazaki, Aichi 444-8585, Japan}

\author{Kazutaka G. Nakamura}
\email{nakamura@msl.titech.ac.jp}
\affiliation{Laboratory for Materials and Structures, Institute of Innovative Research, Tokyo Institute of Technology, 4259 Nagatsuta, Yokohama 226-8503, Japan}
\date{\today}
\begin{abstract}
The coherent control of optical phonons has been experimentally demonstrated in various physical systems. While the transient dynamics for optical phonons can be explained by phenomenological models, 
the coherent control experiment cannot be explained due to the quantum interference. Here, we theoretically propose the generation and detection processes of the optical phonons 
and experimentally confirm our theoretical model using the diamond optical phonon by the double-pump-probe type experiment.
\end{abstract}
\maketitle

\section*{Introduction}
Coherent control was originally developed for controlling chemical reactions using coherent two-photon processes, in which an electronic excited state was used as an intermediary to assist the chemical reaction to the electronic ground-state potential surface.\cite{Tannor1985, Brumer1986, Tannor1986} Coherent control has been performed for other physical properties, for example, electronic, vibrational and rotational states of atoms and molecules\cite{Unanyan1998, Weinacht1999, Weinacht1999, Meshulach1998, Ohmori2006, Katsuki2006, Branderhorst2008, Ohmori2009, Noguchi2014, Higgins2017} and excitons, spins, and phonons in the solid state\cite{Weiner1991, Dekorsy1993C, Hase1996, Bonadeo1998, Wehner1998, Dudovich2003, Dutt2007, Press2008, Takahashi2009, Kozak2013, Katsuki2013, Katsuki2013B, Kono2013, Tahara2014, Hayashi2014, Hase2015, Pingault2017} and the superconducting electrical circuits.\cite{Nakamura1999,Oliver2006,Houck2011} 

The coherent control of optical phonons was first demonstrated in the molecular crystals at cryogenic temperature using multiple femtosecond pulses.\cite{Weiner1991} 
This was well explained by an impulsive stimulated Raman scattering (ISRS) mechanism to generate the coherent optical phonon. However, the similar coherent control experiments with double femtosecond pulses were performed on semimetal films\cite{Hase1996} to be explained by a displacive excitation mechanism.\cite{Cheng1991,Zeiger1992} To understand the unified generation mechanism of coherent phonons, the microscopic theory based on quantum dynamics is required.\cite{Reiter2011, Nakamura2015, Wigger2016, Watanabe2017} However, in the coherent control experiment, the amplitude and phase dependences have been not yet understood from these microscopic theories. 

The aim of this paper is to theoretically propose the unified process included the generation and detection of the coherent optical phonons from quantum dynamics under a two-electronic-level and 
a displaced harmonic oscillator model under the off-resonant condition to extend Ref.~\onlinecite{Nakamura2015}. The effect of detuning and the control scheme are discussed. 
In addition, we demonstrate the coherent control of $40$ THz optical phonons in diamond using a pair of sub-$10$-fs optical pulses by the ISRS process since the band gap of diamond is 
well above the energy of a commonly used femtosecond laser pulse. 

As an application to quantum information technology, diamond is expected to be applied to quantum memory using the nitrogen or silicon-vacancy center in diamond~\cite{Dutt2007, Kagami2010, Fuchs2011, Lukin2017} and the 
optical phonon \cite{Lee2011, Lee2011B, England2013, England2015} since the high-purity material is available and it is working at room temperature. On the other hand, the phonon property of diamond 
has been discussed in the context of photophysics.~\cite{Grimsditch1978, Zaitsev2001, Ishioka2006, Nakamura2016} To understand the coherence of the optical phonon fundamentally and practically, our coherent control scheme 
might be helpful.

\section*{Results}
\subsection*{Two-electric-level coherent-phonon generation and detection model}
It was shown that the generation and detection processes of coherent phonons can be described by the two-band 
density matrix formalism with the optical response function.\cite{Kayanuma2017} 
It was assumed that the band-gap energy is modulated by the coherent oscillation of the optical phonon due to the deformation potential interaction.
In the case of excitation to the transparent wavelength region in diamond treated in the present work, 
we may adopt a much simplified version of this theory.

Let us consider a two-level system for the electronic state coupled with a harmonic oscillator.\cite{Nakamura2015, Mukamel1995}
The Hamiltonian is given by
\begin{eqnarray}
H_0&=&H_g|g\rangle\langle g|+ (\epsilon+ H_e) |e\rangle\langle e|,\nonumber\\
H_g&=&\hbar\omega b^\dagger b,\nonumber\\
H_e&=&\hbar\omega b^\dagger b + \alpha \hbar\omega \left(b+b^\dagger\right),
\end{eqnarray}
where the state vector $\ketg$ refers to the electronic ground state of the crystal, and $ |e\rangle $ refers to the electronic excited state with the excitation energy $\epsilon$. 
In the case of diamond, $|e \rangle$ corresponds representatively 
to the electronic states above the direct band gap, and $\epsilon$ is approximately equal to the direct band gap 
energy $7.3$ eV.\cite{Saslow1966, Milden2013}

The Hamiltonian $H_g$ and $H_e$ are the phonon Hamiltonians in the subspaces $\ketg$ and $ |e\rangle $. 
Here we have introduced the annihilation and the creation operator $b$ and $b^\dagger$ 
for the interaction mode\cite{Kayanuma2017} which is defined as a linear combination of the normal modes 
lying close to the $\Gamma$ point in the Brillouin zone. 
Because of the phase-matching condition, the wave vector of the phonon 
is equal to the wave vector of the incident photon modulated by the by refractive index of the crystal. In the case of 
coherent phonons, the incident pulse is decomposed into a linear combination of plane waves around the central 
mode. Therefore, the wave vector of the coherent phonon is also distributed over a small region around 
$\Gamma$ point in the Brillouin zone. The dispersion of the optical 
phonon energy near the $\Gamma$ point is neglected, and we set the energy of the interaction mode $\hbar\omega$ 
is equal to the optical phonon energy at $\Gamma$ point. 
In the case of diamond, $\omega$ is evaluated as $\sim 2\pi\times 40$THz.\cite{Ishioka2006, Nakamura2016} 
Note that the transverse and longitudinal optical phonon cannot be distinguished at $\Gamma$ point because of a non-polar material. 
The dimensionless coupling constant is denoted by $\alpha $. 
In the bulk crystal, the Huang--Rhys factor $\alpha^2$ is considered to be small; $\alpha^2 \ll 1$. 
For simplicity, we consider a four-state model with two phonon states for each electronic state: $ | g, 0 \rangle $ and $ | g, 1 \rangle $ 
for phonon Fock states with $n=0$ and 1 in the electronic ground state and $ | e, 0 \rangle $ and $ | e, 1 \rangle $ for those in the electronic excited state.
The interaction Hamiltonian with the optical pulse is given by
\begin{equation}
H_I= \mu E_0 f(t) (e^{-i \Omega t} |e\rangle\langle g| + e^{i \Omega t} |g\rangle\langle e|),
\label{int}
\end{equation}
in which $\mu$ is the transition dipole moment from $\ketg$ to $ |e\rangle $, and $E_0$, $\Omega$ and $f(t)$ are the strength, central frequency and temporal profile of the electric field of the pump pulse, respectively. 
The time evolution of the density operator was obtained by solving the quantum Liouville equation using a perturbative expansion in the lowest order. 

We restricted the well separated pulses for the two pump pulses (pump 1 and pump 2) and the probe pulse. Then the generation and detection processes were separately treated, which corresponded to a doorway-window picture in nonlinear spectroscopy.\cite{Mukamel1995, Yan1991} When pump 1 and 2 were well separated, the excitation of the optical phonons occurred with each pulse. The pathway of electronic excitation by pump 1 and the deexcitation by pump 2 was not allowed for the off-resonant condition.
We set the initial state in $| g, 0 \rangle$, then $\rho (- \infty) = | g, 0\rangle \langle g, 0|$. 
This was reasonable for the diamond case, because the population number in the $n=1$ state was approximately $0.005$ at $300$ K. 
There were four Liouville pathways for the exciting phonon polarization: $| g, 0 \rangle \langle g, 0| \rightarrow | e, 0 \rangle \langle g, 0| \rightarrow | g, 1 \rangle \langle g, 0| $, $| g, 0 \rangle \langle g, 0| \rightarrow | e, 1 \rangle \langle g, 0| \rightarrow | g, 1 \rangle \langle g, 0| $ and Hermitian conjugates for each pump pulse.

The density operator for the excitation by pump 1 $\rho^{(2)}_{1}(t)$ was obtained as
\begin{eqnarray}
\rho^{(2)}_{1}(t) & = & \alpha \frac{ \mu^2 |E_1|^2}{\hbar^2} e^{-i\omega t}\int_{-\infty}^\infty dt^{\prime} \int_{-\infty}^{t^{\prime} }dt^{\prime \prime} f_1(t^{\prime})f_1(t^{\prime \prime}) \nonumber \\ 
& \times & \left(e^{i\omega t^{\prime}} - e^{i\omega t^{\prime \prime}}\right) e^{-i\Delta (t^{\prime}-t^{\prime \prime})} |g,1\rangle \langle g,0|,
\label{int2}
\end{eqnarray}
where $|E_1|$ is the strength of the electric field of pump 1, $\Delta \equiv \epsilon/\hbar - \Omega$ is the detuning, and we assumed that $t$ is well after the passage of the pump pulse. 

In the case for far off-resonance excitation, the density matrix can be evaluated as follows. For simplicity, we assume a Gaussian pulse with pulse-width $\sigma$, 
\begin{equation}
f(t)=\frac{1}{\sqrt{\pi}\sigma \Omega}\exp \left( -\frac{t^2}{\sigma^2} \right),
\label{gausspulse}
\end{equation}
where $\Omega$ is used to make the normalization factor, $\sqrt{\pi} \sigma \Omega$, dimensionless. 
Following the calculations to Ref.~\onlinecite{Nakamura2015}, we obtain 
\begin{equation}
\rho^{(2)}_{1}(t)=i\alpha\frac{\mu^2 |E_1|^2}{\hbar^2}e^{-i\omega t}\frac{\omega}{\sqrt{2}\sigma \Omega^2 \Delta^2}
e^{-\sigma^2\omega^2/8}|g, 1\rangle\langle g, 0|.
\end{equation}
The density operator for the excitation by pump 2, $\rho^{(2)}_{2}(t)$, was obtained in a similar calculation, and we obtain the density operator $\rho^{(2)} (t)=\rho^{(2)}_{1}(t)+\rho^{(2)}_{2}(t)$
\begin{eqnarray}
\rho^{(2)} (t) &=&i\alpha\frac{\mu^2}{\hbar^2}\frac{\omega}{\sqrt{2}\sigma \Omega^2 \Delta^2}
e^{-\sigma^2\omega^2/8}\nonumber\\
&\times&\left( |E_1|^2 e^{-i\omega t}+ |E_2|^2 e^{-i\omega (t-\tau)}\right)|g, 1\rangle\langle g, 0| \nonumber\\
&=&i A \left( |E_1|^2 + |E_2|^2 e^{i\omega \tau}\right) \times e^{-i \omega t} |g, 1\rangle\langle g, 0|,
\end{eqnarray}
where 
\begin{eqnarray}
A \equiv \alpha\frac{\mu^2 }{\hbar^2}\frac{\omega}{\sqrt{2}\sigma \Omega^2 \Delta^2} e^{-\sigma^2\omega^2/8},
\end{eqnarray}
$\tau$ is the delay between the pump 1 and the pump 2, which is called the pump-pump delay, and $E_2$ is the electric field strength of pump 2.

The coherent phonon dynamics can be investigated by calculating the mean value of the phonon coordinate $\langle Q (t) \rangle = \mathrm{Tr}\{ Q \rho ^{(2)} (t) \}$, where $Q \equiv \sqrt{\hbar/2\omega}\left(b+b^\dag\right)$ and $\mathrm{Tr}$ indicate the trace. By considering the Hermitian conjugated paths, we obtain
\begin{eqnarray}
\langle Q (t) \rangle &= & \sqrt{\frac{\hbar}{2\omega}}\alpha\frac{\mu^2 }{\hbar^2} \frac{\omega}{\sqrt{2}\sigma \Omega^2 \Delta^2} e^{-\sigma^2\omega^2/8} \nonumber\\
& \times & \big\{ |E_1|^2 \sin (\omega t) + |E_2|^2 \sin (\omega (t-\tau) \big\} \nonumber\\
&= & \sqrt{\frac{\hbar}{2\omega}} A \big\{ |E_1|^2 \sin (\omega t) + |E_2|^2 \sin (\omega (t-\tau) \big\}.
\label{Q}
\end{eqnarray}

Therefore, the amplitude of the phonon oscillation controlled by the two short pulses is expressed by a sum of the two sinusoidal functions. The phonon amplitude is enhanced 
by two times or canceled when the pump delay matches an integer or half-integer multiple of the vibrational period through constructive or destructive interference, respectively, at the $|E_1|=|E_2|$ condition.
Note that the amplitude of oscillation is inversely proportional to the square of detuning from the excited state.

When the heterodyne detection of the transmitted light is investigated, the detection intensity $I_h(t)$ should be
\begin{eqnarray}
I_h(t) = \Omega l \times {\rm Im} \left[ E_3^*(t) P_s (t) \right],
\end{eqnarray}
where $E_3(t)$ is the strength of the electronic field of the probe pulse, $P_s(t)$ is the polarization at time t, and $l$ is the thickness of the sample.\cite{Mukamel1995} 
The probe pulse irradiates the sample at pump-probe delay $t_p$.
There are eight Liouville pathways for the exciting phonon polarization: 
\begin{description}
\item[path 1] $| g, 1 \rangle \langle g, 0| \rightarrow | e, 1 \rangle \langle g, 0| \rightarrow | g, 0 \rangle \langle g, 0| $,
\item[path 2] $| g, 1 \rangle \langle g, 0| \rightarrow | e, 0 \rangle \langle g, 0| \rightarrow | g, 0 \rangle \langle g, 0| $,
\item[path 3] $| g, 1 \rangle \langle g, 0| \rightarrow | g, 1 \rangle \langle e, 1| \rightarrow | g, 1 \rangle \langle g, 1| $,
\item[path 4] $| g, 1 \rangle \langle g, 0| \rightarrow | g, 1 \rangle \langle e, 0| \rightarrow | g, 1 \rangle \langle g, 1| $,
\end{description}
and their Hermitian conjugates.

We obtain $\rho^{(3)}_1 (t^{\prime})$ for the path 1 as
\begin{eqnarray}
\rho^{(3)}_1 (t^{\prime}) &=& i A (|E_1|^2+ |E_2|^2 e^{i \omega \tau}) \frac{i\mu}{\hbar} E_3 e^{-i \omega t_{p}} \nonumber\\
&\times& \int_{-\infty}^{t^{\prime}} dt^{\prime \prime} f_{3}(t^{\prime \prime}) e^{-i\omega t^{\prime \prime}}  e^{-i \Omega t^{\prime \prime}} e^{-i(\epsilon + \hbar \omega) (t^{\prime}-t^{\prime \prime})/\hbar } \nonumber\\
&\times & |e,1 \rangle \langle g, 0 |, \nonumber\\
\end{eqnarray}
where $f_3(t^{\prime \prime})$ is the Gaussian pulse and $t_{p}$ is the pump-probe delay. 
For the polarization operator $P^{op} = \mu |g  \rangle \langle e| + \mu^* |e  \rangle \langle g| $, the complex polarization at time $t$ is given by $P(t) = {\rm Tr} \{ \rho^{(3)} (t) P^{op} \}$. Then the polarization ($P_1(t^{\prime})$) for the path 1 is given by
\begin{eqnarray}
P_1 (t^{\prime}) &=& \alpha A (|E_1|^2+ |E_2|^2 e^{i \omega \tau}) \frac{\mu^2}{\hbar} E_3 e^{-i \omega t_{p}} \nonumber\\
&\times& \int_{-\infty}^{t^{\prime}} dt^{\prime \prime} f_{3}(t^{\prime \prime}) e^{-i\omega t^{\prime \prime}}  e^{-i \Omega t^{\prime \prime}} e^{-i(\epsilon + \hbar \omega) (t^{\prime}-t^{\prime \prime})/\hbar }, \nonumber\\
\end{eqnarray}
and the time-integrated intensity, $I_1 (t_p) $, of the product between the probe light and polarization is
\begin{eqnarray}
I_1(t_p) &=& \int^{\infty}_{-\infty} E_3 f_3^* (t^{\prime}) P_1(t^{\prime}) dt^{\prime} \nonumber\\
&=& \alpha A (|E_1|^2+ |E_2|^2 e^{i \omega \tau}) \frac{\mu^2}{\hbar} |E_3|^2 e^{-i \omega t_{p}} \nonumber\\
&\times& \int^{\infty}_{-\infty} dt^{\prime}
\int_{-\infty}^{t^{\prime}} dt^{\prime \prime}f_3(t^{\prime}) e^{i \Omega t^{\prime}} f_{3}(t^{\prime \prime}) e^{-i\omega t^{\prime \prime}} \nonumber\\
& \times & e^{-i \Omega t^{\prime \prime}} e^{-i(\epsilon + \hbar \omega) (t^{\prime}-t^{\prime \prime})/\hbar }. 
\end{eqnarray}
Using the Gaussian pulse (\ref{gausspulse}), we obtain
\begin{eqnarray}
I_1(t_p) &=& \alpha A (|E_1|^2+ |E_2|^2 e^{i \omega \tau}) \frac{\mu^2}{\hbar \pi \sigma^2 \Omega^2} |E_3|^2 e^{-i \omega t_{p}} \nonumber\\
&\times& \int^{\infty}_{-\infty} e^{-2 s^2/ \sigma^2} e^{ -i \omega s}ds
\int_{0}^{\infty} du e^{-u^2/(2 \sigma^2)} e^{i (\Delta -\omega/2) u} \nonumber\\
& \approx & (|E_1|^2+ |E_2|^2 e^{i \omega \tau}) \frac{\mu^2}{\hbar \pi \sigma^2 \Omega^2} |E_3|^2 \nonumber\\
& \times & e^{-i \omega t_{p}} \frac{\sqrt{\pi}}{\sqrt{2} \sigma} e^{-\sigma^2 \omega^2 /8} 
\frac{i \alpha A}{\Delta -\omega/2} \nonumber\\
& = & i B (|E_1|^2+ |E_2|^2 e^{i \omega \tau}) e^{-i \omega t_{p}} \frac{1}{\Delta -\omega/2},
\end{eqnarray}
where $s \equiv (t^{\prime}+t^{\prime \prime})/2, u \equiv t^{\prime} - t^{\prime \prime}$, and 
\begin{eqnarray}
B \equiv \frac{\sqrt{\pi} \omega \alpha^2 \mu^2}{2 \sigma^2 \Omega^4 \Delta^2 \hbar^3} |E_3|^2 e^{- \omega^2 \sigma^2 /4}.
\end{eqnarray}
A similar calculation shows that $I_4 (t_p) = I_1(t_p)$ and 
\begin{eqnarray}
I_2(t_p) & = & I_3(t_p) \nonumber\\
& = & -i B (|E_1|^2+ |E_2|^2 e^{i \omega \tau}) e^{-i \omega t_{p}} \frac{1}{\Delta +\omega/2}.
\end{eqnarray}
Then we find
\begin{eqnarray}
I(t_p) & = & \sum I_i(t_p) \nonumber\\
&=& (|E_1|^2+ |E_2|^2 e^{i \omega \tau}) e^{-i \omega t_{p}} \frac{2iB \omega}{\Delta^2}.
\end{eqnarray}
By considering the Hermitian conjugate paths, the time-integrated intensity of the heterodyne detection $I_h(\tau, t_p)$ is given by
\begin{eqnarray}
I_h(\tau, t_p) & = & \frac{\sqrt{\pi} \omega^2 \alpha^2 \mu^2}{ \sigma^2 \Omega^4 \Delta^4 \hbar^3} |E_3|^2 e^{- \omega^2 \sigma^2 /4} \nonumber\\
& \times & \{ |E_1|^2 \sin(\omega t_p) + |E_2|^2 \sin(\omega ( t_p- \tau)) \} \nonumber \\
& = & C (\tau) \sin(\omega t_p - \Theta (\tau)), 
\label{finaleq}
\end{eqnarray}
where 
\begin{eqnarray}
C (\tau) & = & \frac{\sqrt{\pi} \omega^2 \alpha^2 \mu^2}{ \sigma^2 \Omega^4 \Delta^4 \hbar^3} |E_3|^2 e^{- \omega^2 \sigma^2 /4} \nonumber \\
& \times & |E_1|^2 \sqrt{1 + 2 \frac{|E_2|^2}{|E_1|^2} \cos (\omega \tau) + \left( \frac{|E_2|^2}{|E_1|^2} \right)^2}, \label{cp_amp} \\
\Theta (\tau) & = & \arctan \left( \frac{\sin (\omega \tau)}{\cos (\omega \tau) + \frac{|E_2|^2}{|E_1|^2}} \right). \label{cp_phase}
\end{eqnarray}

The present model calculation clearly shows that the response of the transmitted light intensity measured with heterodyne detection exhibits the same dependence on the pump-pump delay 
as that of the mean value of the phonon coordinate. 

\subsection*{Single-pump transmission experiment} 
\begin{figure}[thb]
\centering
\includegraphics[width=8.0 cm]{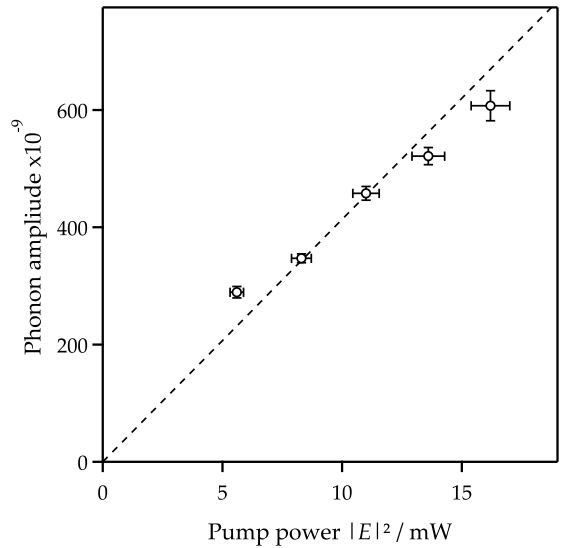}
\caption{Transient transmittance change of diamond. The oscillation (a) is excited by the first pump pulse only, and the oscillation (b) is excited by the second pump pulse only. It is noted that the baseline in our previous experiment,\cite{Nakamura2016} seems to more flat compared to the present one. This is because the experimental data shown in Ref.~\onlinecite{Nakamura2016} has been already 
subtracted by the smoothing curve of the obtained experimental data to easily analyze this. }
\label{singlepulse}
\end{figure}

In the followings, the experimental detection of the optical phonons was performed using a pump-probe type transient transmittance measurement with femtosecond pump pulses, 
see the details in Methods. The transient transmittance change induced by only pump 1 or pump 2 were measured in Fig. \ref{singlepulse} (a) and (b), respectively, against the pump-probe delay $t_{p}$ 
between $-200$ and $1000$ fs. 
It is noted that time zero was set at the time when pump 2 irradiates the sample; 
the minimum portion of the sharp response. 
After the sharp peak, which arose from the nonlinear response for overlapped pump and probe pulses, 
there was a modulation caused by the coherent optical phonons in diamond. The oscillation period was $25.1 \pm 0.03$ fs (frequency of $39.9 \pm 0.05$ THz). 
The coherent oscillation in the transmitted pulse intensity arising from the optical phonons was the same as that obtained by the reflection experiments.\cite{Ishioka2006, Nakamura2016}
\begin{figure}[thb]
\centering
\includegraphics[width=8.0 cm]{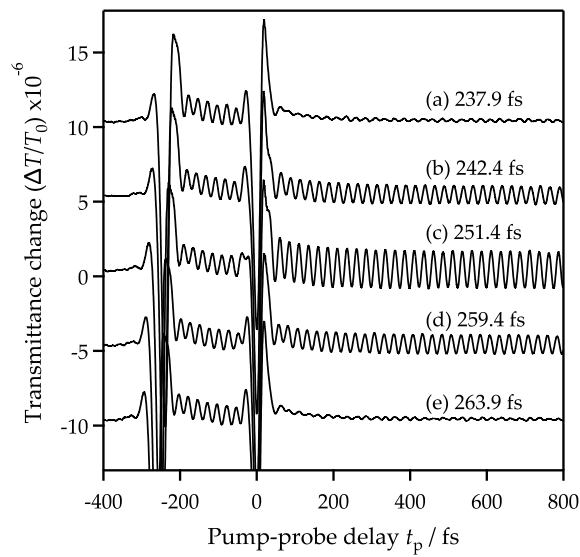}
\caption{The pump power dependence of the amplitude of the 40 THz oscillation. It is noted that the statistical average was $4,800$ signals.}
\label{power}
\end{figure}

To verify the theoretical treatment on the detection process in the previous section, Figure \ref{power} shows a pump laser power dependence of the oscillation amplitude of the $40$ THz oscillation 
in the transmittance spectrum. According to Eq. (\ref{finaleq}) with $|E_2|^2 = 0$, the oscillation amplitude is a linear dependence on the pump laser intensity $|E_1|^2$.
This is well agreement with our experiment data. It is noted that the deviation between our prediction and experimental data has not yet been identified such as 
the laser power and measurement-setup stability.

\subsection*{Coherent control experiment}
\begin{figure}[thb]
\centering
\includegraphics[width=8.5 cm]{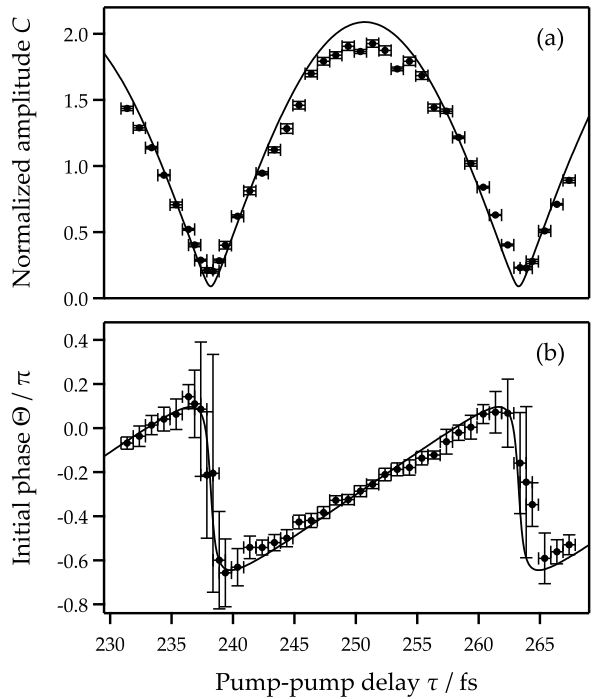}
\caption{Transient transmittance change along the pump-probe delay (between pump 2 and the probe) for several pump-pump delays ($\tau$ between pump 1 and pump 2): at $237.9$ fs (a), $242.4$ fs (b), 
$251.4$ fs (c), $259.4$ fs (d), and $263.9$ fs (e). In this figure, the time zero is set at the timing of the pump 2 irradiation. Each curve is plotted with vertical offsets.}
\label{doublepulse}
\end{figure}

\begin{figure}[thb]
\centering
\includegraphics[width=8.5 cm]{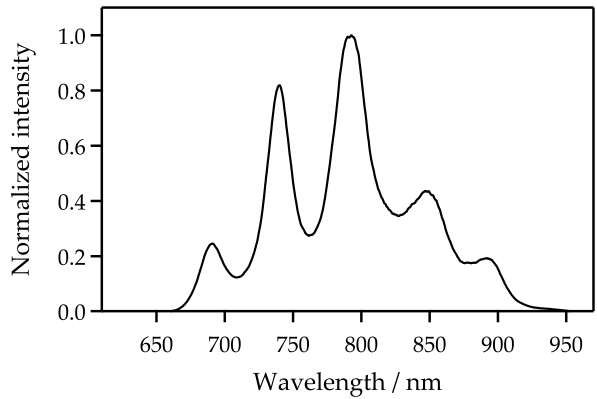}
\caption{The amplitude (a) and phase (b) of the controlled oscillation after pump 2 against the pump-pump delay $\tau$. The amplitude is normalized using that obtained after excitation after only pump 1; oscillation between the pump 1 and pump 2 irradiation timing. Solid circles are the experimental data and the solid curves are obtained by calculation using Eqs. (\ref{cp_amp}) and (\ref{cp_phase}) with $|E_2|^2/|E_1|^2=1.09$, $\omega = 2 \pi \times 39.9$ THz, and the offset initial phase $-0.29 \pi$.}
\label{amplitude}
\end{figure}

Figure \ref{doublepulse} shows typical examples of the transient transmittance changes induced by the pair of pump pulses (pump 1 and 2).
Pump 1 induces a coherent oscillation in the transmission intensity with a frequency of $39.9 \pm 0.05$ THz. 
This oscillation amplitude was controlled by pump 2. It was reduced at $\tau =237.9$ fs, enhanced at $\tau =251.4$ fs, and reduced again at $\tau =263.9$ fs.  
The oscillation amplitude after pump 2 was obtained by fitting a sinusoidal wave in the range of the pump 2-probe delay between $200$ and $650$ fs. 
In this time range, the decrease of the amplitude was negligibly small. The obtained amplitude was plotted along the separation time $\tau$ in Fig. \ref{amplitude}(a). 
We also estimated the initial phase of the oscillation after pump 2 by extrapolation of the fitted sinusoidal function at the timing of the pump 2 irradiation. 
The estimated initial phase is plotted in Fig. \ref{amplitude} (b). Measurement error to define the timing of the pump 2 irradiation was approximately $\pm 0.5 $ fs, 
which corresponds to $\pm 0.04 \pi$ for the phase of the $39.9$ THz oscillation.

In Fig. \ref{amplitude}, the phonon amplitude was normalized by the phonon amplitude excited by pump 1, which was observed at the pump 2-probe delay between $-270$ and $0$ fs. 
The phonon amplitude was enhanced almost twice at $\tau =251.4$ fs and diminished at $\tau =237.9$ fs by the constructive and destructive interference of phonon states.

The energy of the optical pulses (around $1.5$ eV) was well below the direct band gap ($7.3$ eV) of diamond.\cite{Saslow1966, Milden2013} 
Therefore, the coherent optical phonons should be excited by the ISRS process\cite{Nakamura2015} at 
an off-resonant condition. The initial phonon state at room temperature was well expressed with the $n=0$ state, because the phonon energy ($39.9$ THz $\simeq 135$ meV) was higher than 
the thermal energy ($\sim 25$ meV) and the population ratio between the ground and excited state of the optical phonon was $0.005$.
The coherent control of the phonon amplitude shown in Fig. \ref{amplitude} was calculated using our proposed model (\ref{finaleq}). In this experiment, the intensity ratio between the light 
intensities of pump 1 and 2 was $|E_2|^2/|E_1|^2=1.09$ and approximately agreed with the amplitude ratio of the coherent phonon, $\Delta T_2 / \Delta T_1 = 0.99 \pm 0.03$ according 
to the single-pump experiment in Fig.~\ref{singlepulse}. 
The transmission intensity change depending on the pump-pump delay $\tau$
was calculated using Eq. (\ref{finaleq}) and the frequency ($\omega = 2 \pi \times 39.9$ THz) for the optical phonon of diamond. The calculated result is shown in Fig. \ref{amplitude}(a), 
where the intensity was normalized to that induced by a single pump pulse (pulse 1). The initial phase of the oscillation after irradiation of pump 2 was also obtained from the calculation 
and shown in Fig. \ref{amplitude}(b). Our proposed model for the coherent control of the optical phonons reasonably represents the experimental data. 
It is noted that the initial phase shift $-0.29 \pi$, which corresponds to $7.3$ fs, does not be explained in our proposed model. 
This may come from the calibration error of the pump-pump delay and the unknown mechanism of the phonon generation timing. 

Although the phonon amplitude was not directly observed, the observed transmission intensity change (\ref{finaleq}) by heterodyne detection 
has the same pump-pump-delay dependence (\ref{Q}) for the phonon amplitude. If the Huang-Rhys factor and the transition dipole were obtained from other experiments or calculations, 
the phonon amplitude can be estimated from the transient transmittance change.
The phonon coherence induced by the pump pulse was detected by heterodyne detection. The coherent control by two separated pulses was expressed as the phonon coherences induced 
by each pump pulse (pump 1 and pump 2) interfering with each other. 

\section*{Discussions}
In summary, we investigated the coherent control of the optical phonons using a pair of optical pulses with two electronic levels and two harmonic phonon levels. The calculations showed 
that the controlled phonon amplitude and transmission intensity can be expressed by the sum of two sinusoidal functions. Furthermore, we demonstrated a coherent control of the optical 
phonons in a single crystal diamond. We used a pump and probe protocol and the change in the transmitted light intensity was determined with 
heterodyne detection. The phonon amplitude was coherently controlled by changing the the pump-pump delay from $230$ fs to $270$ fs. The control scheme 
was well explained by our theoretical generation and detection model with the interference between the two phonon states excited by each pump pulse.

The wave packet dynamics of the coherent optical phonons is only measured in the transmission intensity change. Therefore, the amplitude and the phase of the 
wave packet cannot be individually controlled. To reproduce the wave packet of the coherent phonon, the transmittance and reflectivity changes should be 
simultaneously measured. According to the Kramers-Kr\"onig relation, the transient complex dielectric constant can be measured. The optical phonon amplitude 
is also measured by combining to the Raman spectroscopy. Furthermore, there still are open questions on the nonlinear response of the optical phonon. The coherent 
control of the optical phonon around $\tau \sim 0$ except for the pulse overlap region might give an insight on generation and detection processes 
for the optical phonon. 

\section*{Methods}
\begin{figure}[thb]
\centering
\includegraphics[width=8.5 cm]{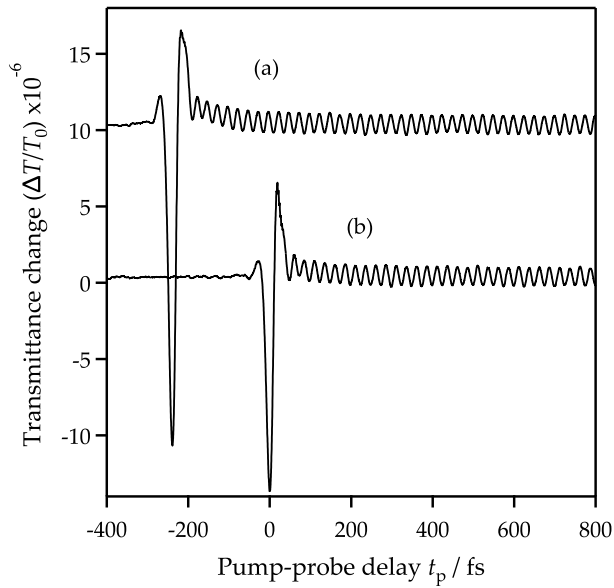}
\caption{The measured spectrum property of ultrafast laser.}
\label{laser}
\end{figure}

The experimental setup has been described in the previous paper\cite{Nakamura2016} in addition to that the optical pulses were generated by using a home-made 
Michelson-type interferometer.\cite{Hayashi2014} While the transient reflectivity change was measured in Ref.~\onlinecite{Nakamura2016}, in this experiment, 
the transient transmittance change is measured. According to Fig.~\ref{laser}, the ultrafast laser conditions measured immediately behind the output port 
were a maximum-intensity wavelength of $792$ nm with a estimated pulse width of $8.2$ fs as full width at half maximum under the assumption of the transform-limited pulse. 
This also has a repetition rate of $75$ MHz. 
To reduce the statistical error, $3,200$ signals were averaged and taken as the measured value. By converting the temporal motion of the scan delay unit to the pump-probe delay, 
the temporal evolution of the change in the transmitted light intensity, $\Delta T(t_p)/T_0$, was obtained. Here we used the heterodyne detection technique.
The powers of the pump 1 and 2 and the probe were $19.1$ mW, $20.8$ mW, and $3.0$ mW, respectively. 
The sample used was a single crystal of diamond with a $[100]$ crystal plane, which was fabricated by chemical vapor deposition 
and obtained from EPD corporation. The type of diamond was intermediate between Ib and IIa and its size was $5 \times 5$ mm$^2$ and $0.7$ mm thick. 
The polarization of the pump and probe pulses were set along the $[110]$ and $[-110]$ axes, respectively. 

\begin{acknowledgements}
The authors thank Yuki Okuda and Mayuko Kato for technical assistance.
This work was supported in part by Core Research for Evolutional Science and Technology (CREST) of the Japan Science and Technology Agency (JST), 
JST ERATO (Grant No. JPMJER1601), JSPS KAKENHI Grant Numbers 15H02103, 15K13377, 16K05410, 17K19051, and 17H02797, 
Collaborative Research Project of Laboratory for Materials and Structures, Institute of Innovative Research, Tokyo Institute of Technology, and Joint Studies Program of Institute for Molecular Science.
\end{acknowledgements}

\subsection*{Author Contribution}
Y.S. and K.G.N. conducted the project.
H.S. and R.T. measured the experimental data. 
H.S., R.T., and Y.O. analyzed the experimental data under the guidance of F.M. and Y.S..
K.G.N. calculated the theoretical model with the help of Y.K. and Y.S..
Y.S. and K.G.N. mainly wrote the manuscript.
All authors discussed the results and commented on the manuscript.
\subsection*{Competing Interests}
The authors declare no competing interests.


\begin{thebibliography}{99}
\bibitem{Tannor1985} Tannor, D. J. \& Rice, S. A. Control of selectivity of chemical reaction via control of wave packet evolution. {\it J. Chem. Phys.} \textbf{83}, 5013 (1985). 
\bibitem{Brumer1986} Brumer, P. \& Shapiro, M. Control of unimolecular reactions using coherent light. {\it Chem. Phys. Lett.} \textbf{126}, 541 (1986).
\bibitem{Tannor1986} Tannor, D. J., Kosloff, R. \& Rice, S. A. Coherent pulse sequence induce control of selectivity of reactions: exact quantum mechanical calculations. {\it J. Chem. Phys.} \textbf{85}, 5805 (1986). 

\bibitem{Unanyan1998} Unanyan, R., Fleischhauer, M., Shore, B. W., \& Bergmann, K. Robust creation and phase-sensitive probing of superposition states via stimulated Raman adiabatic passage (STIRAP) with degenerated dark states. {\it Opt. Commun.} \textbf{155}, 144 (1998). 
\bibitem{Weinacht1999} Weinacht, T. C., Ahn, J., \& Bucksbaum, P. H. Controlling the shape of a quantum wavefunction. {\it Nature} \textbf{397}, 233 (1999). 
\bibitem{Meshulach1998} Meshulach, D. \& Silberberg, Y. Coherent quantum control of two-photon transitions by a femtosecond laser pulse. {\it Nature} \textbf{396}, 239 (1998). 
\bibitem{Ohmori2006} Ohmori, K., Katsuki, H., Chiba, H., Honda, M., Hagihara, Y., Fujiwara, K., Sato, Y., \& Ueda, K. Real-Time Observation of Phase-Controlled Molecular Wave-Packet Interference. {\it Phys. Rev. Lett.} \textbf{96}, 093002 (2006). 
\bibitem{Katsuki2006} Katsuki, H., Chiba, H., Girard, B., Meier, C., \& Ohmori, K. Visualizing Picometric Quantum Ripples of Ultrafast Wave-Packet Interference. {\it Science} \textbf{311}, 1589 (2006).
\bibitem{Branderhorst2008} Branderhorst, M. P. A., Londero, P., Wasylcyk, P., Brif, C., Kosut, R. L., Rabitz, H., \& Walmsley, I. A. Coherent Control of Decoherence. {\it Science} \textbf{320}, 638 (2008). 
\bibitem{Ohmori2009} Ohmori, K. Wave-Packet and Coherent Control Dynamics. {\it Annu. Rev. Phys. Chem.} \textbf{60}, 487 (2009).
\bibitem{Noguchi2014} Noguchi, A., Shikano, Y., Toyoda, K., \& Urabe, S. Aharonov-Bohm effect in the tunnelling of a quantum rotor in a linear Paul trap. {\it Nat. Commun.} \textbf{5}, 3868 (2014).
\bibitem{Higgins2017} Higgins, G., Pokorny, F., Zhang, C., Bodart, Q., \& Hennrich, M. Coherent Control of a Single Trapped Rydberg Ion. {\it Phys. Rev. Lett.} \textbf{119}, 220501 (2017).

\bibitem{Bonadeo1998} Banadeo, N. H., Erland, J., Gammon, D., Park, D., Katzer, D. S., \& Steel, D. G. Coherent Optical Control of the Quantum State of a Single Quantum Dot. {\it Science} \textbf{282}, 1473 (1998). 
\bibitem{Wehner1998} Wehner, M. U., Ulm, M. H., Chemla, D. S., \& Wegener, M. Coherent Control of Electron-LO-Phonon Scattering in Bulk GaAs. {\it Phys. Rev. Lett.} \textbf{80}, 1992-1995 (1998). 
\bibitem{Dutt2007} Gurudev Dutt, M. V., Childress, L., Jiang, L., Togan, E., Maze, J., Jelezko, F., Zibrov, A. S., Hemmer, P. R., \& Lukin, M. D. Quantum Register Based on Individual Electronic and Nuclear Spin Qubits in Diamond. {\it Science} \textbf{316}, 1312 (2007). 
\bibitem{Press2008} Press, D., Ladd, T. D., Zhang, B., \& Yamamoto, Y. Complete quantum control of a single quantum dot spin using ultrafast optical pulses. {\it Nature} \textbf{456}, 218 (2008). 
\bibitem{Kozak2013} Koz\'{a}k, M., Troj\'{a}nek, F., Gal\'{a}\v{r}, P., Varga, M., Kromka, A., \& Mal\'{y}, P. Coherent phonon dynamics in micro- and nanocrystalline diamond. {\it Opt. Exp.} \textbf{21}, 31521 (2013).
\bibitem{Katsuki2013B} Katsuki, H., Kayanuma, Y., \& Ohmori, K. Optically engineered quantum interference of delocalized wave functions in a bulk solid: The example of solid para-hydrogen. {\it Phys. Rev. B} \textbf{88}, 014507 (2013). 
\bibitem{Tahara2014} Tahara, H. \& Kanemitsu, Y. Dynamical coherent control of photocurrent in bulk GaAs at room temperature. {\it Phys. Rev. B} \textbf{90}, 245203 (2014). 
\bibitem{Pingault2017} Pingault, B., Jarausch, D.-D., Hepp, C., Klintberg, L., Becker, J. N., Markham, M., Becher, C., \& Atat\"{u}re, M. Coherent control of the silicon-vacancy spin in diamond. {\it Nat. Commun.} \textbf{8} 15579 (2017).

\bibitem{Weiner1991} Weiner, A. M., Leaird, D. E., Wiederrecht, G. P., \& Nelson, K. A. Femtosecond multiple-pulse impulsive stimulated Raman scattering spectroscopy. {\it J. Opt. Soc. Am. B} \textbf{8}, 1264 (1991). 
\bibitem{Dekorsy1993C} Dekorsky, T., K\"{u}tt, W., Pfeifer, T., \& Kurz, H. Coherent Control of LO-Phonon Dynamics in Opaque Semiconductors by Femtosecond Laser Pulses. {\it Europhys. Lett.} \textbf{23}, 223 (1993). 
\bibitem{Hase1996} Hase, M., Mizoguchi, K., Harima, H., Nakashima, S., Tani, M., Sakai, K., \& Hangyo, M. Optical control of coherent optical phonons in bismuth films. {\it Appl. Phys. Lett.} \textbf{69}, 2474 (1996).
\bibitem{Dudovich2003} Dudovich, N., Oron, D., \& Silberberg, Y., Single-pulse coherent anti-Stokes Raman spectroscopy in the fingerprint spectral region. {\it J. Chem. Phys.} \textbf{118}, 9208 (2003). 
\bibitem{Takahashi2009} Takahashi, H., Kato, K., Nakano, H., Kitajima, M., Ohmori, K. \& Nakamura, K. G. Optical control and mode selective excitation of coherent phonons in ${\rm YBa_2Cu_3O_{7-\delta}}$.
{\it Solid State Commun.} \textbf{149}, 1955 (2009). 
\bibitem{Katsuki2013} Katsuki, H., Delagnes, J. C., Hosaka, K., Ishioka, K., Chiba, H., Zijlstra, E. S., Garcia, M. E., Takahashi, H., Watanabe, K., Kitajima, M., Matsumoto, Y., Nakamura, K. G., \& Ohmori, K. All-optical control and visualization of ultrafast two-dimensional atomic motions in a single crystal of bismuth. {\it Nat. Commun.} \textbf{4}, 2801 (2013).
\bibitem{Kono2013} Kim, J.-H., Nugraha, A. R. T., Booshehri, L. G., H\'{a}roz, E. H., Sato, K., Sanders, G. D., Yee, K.-J., Lim, Y.-S., Stanton, C. J., Saito, R., \& Kono, J. Coherent phonons in carbon nanotubes and graphene. {\it Chem. Phys.} \textbf{413}, 55 (2013). 
\bibitem{Hayashi2014} Hayashi, S., Kato, K., Norimatsu, K., Hada, M., Kayanuma, Y., \& Nakamura, K. G., Measuring quantum coherence in bulk solids using dual phase-locked optical pulses. {\it Sci. Rep.} \textbf{4}, 4456 (2014).
\bibitem{Hase2015} Hase, M., Fons, P., Mitrofanov, K., Kolobov, A. V., \& Tominaga, J. Femtosecond structural transformation of phase-change materials far from equilibrium monitored by coherent phonons. {\it Nat. Commun.} \textbf{6}, 8367 (2015). 

\bibitem{Nakamura1999} Nakamura, Y., Pashkin, Yu. A., \& Tsai, J. S. Coherent control of macroscopic quantum states in a single-Cooper-pair box. {\it Nature} \textbf{398}, 786 (1999).
\bibitem{Oliver2006} Berns, D. M., Oliver, W. D., Valenzuela, S. O., Shytov, A. V., Berggren, K. K., Levitov, L. S., \& Orlando, T. P. Coherent Quasiclassical Dynamics of a Persistent Current Qubit. {\it Phys. Rev. Lett.} \textbf{97}, 150502 (2006).
\bibitem{Houck2011} Hoffman, A. J., Srinivasan, S. J., Gambetta, J. M., \& Houck, A. A., Coherent control of a superconducting qubit with dynamically tunable qubit-cavity coupling. {\it Phys. Rev. B} \textbf{84}, 184515 (2011).

\bibitem{Cheng1991} Cheng, T. K., Vidal, J., Zeiger, H. J., Dresselhaus, G., Dresselhaus, M. S., \& Ippen, E. P. Mechanism for displacive excitation of coherent phonons in Sb, Bi, Te, and ${\rm Ti_2O_3}$. {\it Appl. Phys. Lett.} \textbf{59}, 1923 (1991). 
\bibitem{Zeiger1992} Zeiger, H. J., Vidal, J., Cheng, T. K., Ippen, E. P., Dresselhaus, G., \& Dresselhaus, M. S. Theory for displacive excitation of coherent phonons. {\it Phys. Rev. B} \textbf{45}, 768 (1992). 

\bibitem{Reiter2011} Reiter, D. E., Wigger, D., Axt, V. M., \& Kuhn, T. Generation and dynamics of phononic cat states after optical excitation of a quantum dot. {\it Phys. Rev. B} \textbf{84}, 195327 (2011). 
\bibitem{Nakamura2015} Nakamura, K. G., Shikano, Y., \& Kayanuma, Y. Influence of pulse width and detuning on coherent phonon generation. {\it Phys. Rev. B} \textbf{92}, 144304 (2015).
\bibitem{Wigger2016} Wigger, D., Gehring, H., Axt, V. M., Reiter, D. E., \& Kuhn, T. Quantum dynamics of optical phonons generated by optical excitation of a quantum dot. {\it J. Comput. Electron.} \textbf{15}, 1158 (2016). 
\bibitem{Watanabe2017} Watanabe, Y., Hino, K., Hase, M., \& Maeshima, N., Quantum Generation Dynamics of Coherent Phonons: Analysis of Transient Fano Resonance. {\it Phys. Rev. B} \textbf{95}, 014301 (2017). 

\bibitem{Kagami2010} Kagami, S., Shikano, Y., \& Asahi, K. Detection and manipulation of single spin of nitrogen vacancy center in diamond toward application of weak measurement. {\it Physica E} {\bf 43}, 761 (2011).
\bibitem{Fuchs2011} Fuchs, G. D., Burkard, G., Klimov, P. V., \& Awschalom, D. D. A quantum memory intrinsic to single nitrogen-vacancy centres in diamond. {\it Nat. Phys.} {\bf 7}, 789 (2011).
\bibitem{Lukin2017} Sukachev, D. D., Sipahigil, A., Nguyen, C. T., Bhaskar, M. K., Evans, R. E., Jelezko, F., \& Lukin, M. D. Silicon-Vacancy Spin Qubit in Diamond: A Quantum Memory Exceeding 10 ms with Single-Shot State Readout. {\it Phys. Rev. Lett.} \textbf{119}, 223602 (2017).
\bibitem{Lee2011} Lee, K. C., Sprague, M. R., Sussman, B. J., Nunn, J., Langford, N. K., Jin, X.-M., Champion, T., Michelberger, P., Reim, K. F., England, D., Jaksch, D., \& Walmsley, I. A. Entangling Macroscopic Diamonds at Room Temperature. {\it Science} \textbf{334}, 1253 (2011).
\bibitem{Lee2011B} Lee, K. C., Sussman, B. J., Sprague, M. R., Michelberger, P., Reim, K. F., Nunn, J., Langford, N. K., Bustard, P. J., Jaksch, D., \& Walmsley, I. A. Macroscopic non-classical states and terahertz quantum processing in room-temperature diamond. {\it Nat. Photon.} \textbf{6}, 41 (2012).
\bibitem{England2013} England, D. G., Bustard, P. J., Nunn, J., Lausten, R., \& Sussman, B. J. From Photons to Phonons and Back: A THz Optical Memory in Diamond. {\it Phys. Rev. Lett.} \textbf{111}, 243601 (2013).
\bibitem{England2015} England, D. G., Fisher, K. A. G., MacLean, J.-P. W., Bustard, P. J., Lausten, R., Resch, K. J., \& Sussman, B. J. Storage and Retrieval of THz-Bandwidth Single Photons Using a Room-Temperature Diamond Quantum Memory. {\it Phys. Rev. Lett.} \textbf{114}, 053602 (2015).

\bibitem{Grimsditch1978} Grimsditch, M. H., Anastassakis, E., \& Cardona, M. Effect of uniaxial stress on the zone-center optical phonon of diamond. {\it Phys. Rev. B} {\bf 18}, 901 (1978).
\bibitem{Zaitsev2001} Zaitsev, A. M. {\it Optical Properties of Diamond: A Data Handbook} (Springer-Verlag, Berlin, 2001).
\bibitem{Ishioka2006} Ishioka, K., Hase, M., Kitajima, M., \& Petek, H. Coherent optical phonons in diamond. {\it Appl. Phys. Lett.} \textbf{89}, 231916 (2006).
\bibitem{Nakamura2016} Nakamura, K. G., Ohya, K., Takahashi, H., Tsuruta, T., Sasaki, H., Uozumi, S., Norimatsu, K., Kitajima, M., Shikano, Y., \& Kayanuma, Y. Spectrally resolved detection in transient-reflectivity measurements of coherent optical phonons in diamond. {\it Phys. Rev. B} \textbf{94}, 024303 (2016).

\bibitem{Kayanuma2017} Kayanuma, Y. \& Nakamura, K. G. Dynamic Jahn-Teller viewpoint for generation mechanism of asymmetric modes of coherent phonons. {\it Phys. Rev. B} {\bf 95}, 104302 (2017).
\bibitem{Mukamel1995} Mukamel, S. \textit{Principles of Nonlinear Optical Spectroscopy} (Oxford University Press, New York, 1995).

\bibitem{Saslow1966} Saslow, W., Bergstresser, T. K., \& Cohen, M. L. Band Structure and Optical Properties of Diamond. {\it Phys. Rev. Lett.} \textbf{16}, 354 (1966).
\bibitem{Milden2013} Milden, R. P. in \textit{Optical Engineering of Diamonds}, edited by Milden, R. P. \& Rabeau, J. R. (Wiley-VCH, Weinheim, 2013), pp. 1-34.
\bibitem{Yan1991} Yan, Y. J. \& Mukamel, S. Pulse shaping and coherent Raman spectroscopy in condensed phases. {\it J. Chem. Phys.} \textbf{94}, 997 (1991).

\end{thebibliography}
\end{document}